\documentclass{article}

\usepackage[utf8]{inputenc} 
\usepackage[T1]{fontenc}    
\usepackage{hyperref}       
\usepackage{url}            
\usepackage{booktabs}       
\usepackage{amsfonts}       
\usepackage{nicefrac}       
\usepackage{microtype}      
\usepackage{graphicx}
\usepackage{natbib}
\usepackage{doi}
\usepackage{amsthm}
\usepackage{amsmath}
\usepackage{amssymb}
\usepackage[ruled,linesnumbered]{algorithm2e}
\usepackage{boldline} 
\usepackage{ctable}
\usepackage{appendix}
\usepackage{authblk}
\usepackage[a4paper, total={6in, 9in}]{geometry}

\providecommand{\keywords}[1]
{
  \textbf{Keywords:} #1
}

\title{A Nonparametric Mixed-Effects Mixture Model for Patterns of Clinical Measurements Associated with COVID-19}

\date{} 					

\author[1]{\large Xiaoran Ma}
\author[2]{Wensheng Guo}
\author[1]{Mengyang Gu}
\author[3]{Len Usvyat}
\author[4,5]{Peter Kotanko}
\author[1]{Yuedong Wang\thanks{Corresponding author: yuedong@pstat.ucsb.edu}}
\affil[1]{ \normalsize Department of Statistics and Applied Probability, University of California, Santa Barbara}
\affil[2]{Department of Biostatistics, Epidemiology and Informatics, University of Pennsylvania}
\affil[3]{Fresenius Medical Care}
\affil[4]{Renal Research Institute}
\affil[5]{Icahn School of Medicine at Mount Sinai}

\begin{document}
\maketitle

\begin{abstract}
\normalsize
	Some patients with COVID-19 show changes in signs and symptoms such as temperature and oxygen saturation days before being positively tested for SARS-CoV-2, while others remain asymptomatic. It is important to identify these subgroups and to understand what biological and clinical predictors are related to these subgroups. This information will provide insights into how the immune system may respond differently to infection and can further be used to identify infected individuals. We propose a flexible nonparametric mixed-effects mixture model that identifies risk factors and classifies patients with biological changes. We model the latent probability of biological changes using a logistic regression model and trajectories in the latent groups using smoothing splines. We developed an EM algorithm to maximize the penalized likelihood for estimating all parameters and mean functions. We evaluate our methods by simulations and apply the proposed model to investigate changes in temperature in a cohort of COVID-19-infected hemodialysis patients.
\end{abstract}


\section{Introduction}

The Coronavirus Disease 2019 (COVID-19) pandemic has had a profound impact on humanity, and the emergence of new variants of severe acute respiratory syndrome coronavirus 2 (SARS‑CoV‑2) challenges healthcare systems worldwide. Much research has examined changes in biological variables for diagnosis and prognosis during and after an infection caused by the SARS‑CoV‑2 \citep{PIMENTEL202099,malik_biomarkers_2021,chaudhuri_trajectories_2022}. Early detection of SARS-CoV-2 infection based on changes in readily available measurements such as temperature and arterial oxygen saturation during the incubation period is crucial for isolating and treating contagious individuals \citep{de_moraes_batista_covid-19_2020,wu_rapid_2020,kukar_covid-19_2021,monaghan_machine_2021}. Indicators of disease severity and prognosis are essential to the clinical management of COVID-19 patients \citep{jiang_towards_2020,malik_biomarkers_2021,gallo_marin_predictors_2021}. Identifying potential changes in an individual is challenging because the clinical presentation of COVID-19 varies greatly from asymptomatic infection to critical illness \citep{harahwa_optimal_2020,souza_clinical_2020,da_rosa_mesquita_clinical_2021}. It is difficult to predict how the disease will manifest itself in an individual. Identifying biological variables, their longitudinal patterns, variations between individuals, and associations with demographic and clinical characteristics would aid the development of a risk-stratified approach to patient care.

\begin{figure}
    \centering
    \includegraphics[scale=0.4]{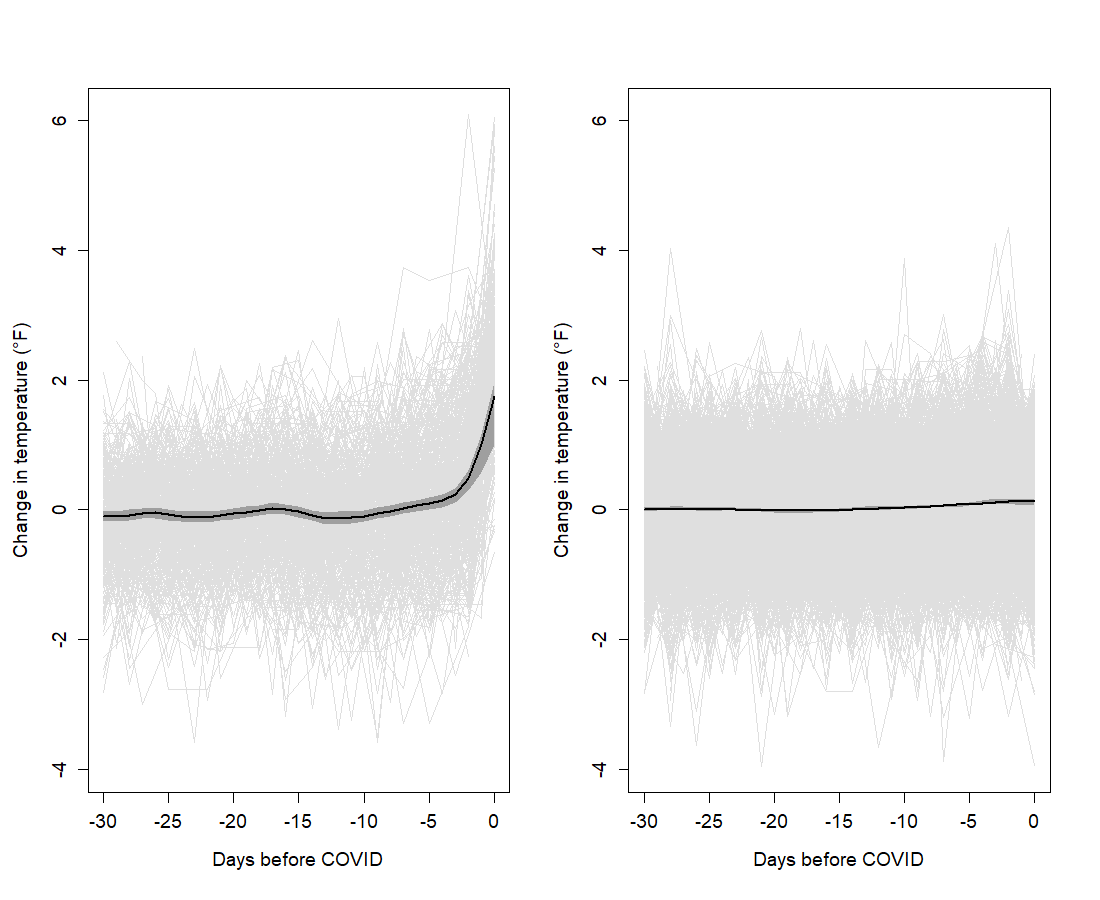}
    \caption{The temperature profiles before confirmation time in 3,293 COVID-19 HD patients. Symptomatic (left) and asymptomatic (right) groups were classified by the method in this paper. Red lines are the estimates of mean functions, and shaded areas are 95\% bootstrap confidence intervals.}
    \label{figPre}
\end{figure}

Nearly 786,000 people in the United States have end-stage renal disease (ESRD). About 488,000 ESRD patients travel to clinics to receive life-sustaining hemodialysis (HD) treatments and cannot shelter in place \citep{noauthor_unites_2020,noauthor_kidney_nodate}. HD patients suffer from a host of comorbidities, such as diabetes and cardiac disease, putting them at an increased risk for complications from COVID-19. In addition, HD patients have reduced responses to SARS-CoV-2 vaccines \citep{simon_haemodialysis_2021}. Thus, there is a pressing need to identify potential coronavirus carriers and develop procedures to curb the spread among HD patients. Numerous studies \citep{bivona_biomarkers_2021,malik_biomarkers_2021,gallo_marin_predictors_2021} focus on the general population but only a few center on HD patients \citep{monaghan_machine_2021}.
The thrice-weekly in-center HD treatments provide results of a large number of clinical and treatment variables that are stored in patients’ electronic health records (EHRs) and thus readily available for analysis. Utilizing EHRs, \cite{chaudhuri_trajectories_2022} observed significant changes in many biological variables due to COVID-19 infection. However, these results estimated at the population level are not directly applicable to individual detection and prediction. For each patient, we compute temperature change as the difference between measured temperatures minus the average temperatures during a period free of COVID-19 infection.  Figure \ref{figPre} presents temperature change profiles before confirmation time in a cohort of COVID-19 HD patients. Since the body temperature of some patients raised a couple of days before being tested positive for COVID-19, these patterns are indicative of COVID-19 infection. Nevertheless, the temperatures of other patients remain relatively unchanged. We observed similar patterns for other biological variables, including pulse rate, systolic blood pressure, interdialytic weight gain, serum levels of albumin and ferritin, and counts of neutrophils and lymphocytes \citep{chaudhuri_trajectories_2022}. 

This paper focuses on estimating changes in biological and clinical indicators in in-center HD patients with COVID-19.  
We have two objectives: (a) clustering patients into two groups: symptomatic and asymptomatic; and (b) associating the group probability with comorbidities and demographic/clinical characteristics. Existing methods for clustering longitudinal and functional data (see  \cite{bouveyron_model-based_2014} and \cite{jacques_functional_2014} for reviews) do not apply since they were proposed only for objective (a).
We proposed a novel Nonparametric Mixed-Effects Mixture (NMEM) model to fulfill both objectives (a) and (b). We model the probability of a latent group label using a logistic regression model and the response variable when the latent group label is given using a nonparametric mixed-effects model.   \cite{joo_clustering_2009} considered a similar model with P-spline fixed effects only for the characteristics of urban groundwater recharge. \cite{lu_finite_2012} studied a Bayesian model with B-spline fixed effects and random intercepts for
the treatment effect of heroin use. 
The main contributions of this paper are as follows: (1) modeling both the group mean and subject deviation trajectories using smoothing splines, which makes the proposed NMEM model more flexible than those in \cite{joo_clustering_2009} and \cite{lu_finite_2012}. Our method extends that in \cite{ma_penalized_2008} by allowing the probability to depend on covariates; (2) introducing an $L_1$ regularization method for variable selection, which \cite{joo_clustering_2009} and \cite{lu_finite_2012} did not consider; and (3) investigating changes in temperature in a cohort of COVID-19-infected hemodialysis patients using the proposed method. We note that the proposed method is general, which is not limited to the application illustrated in this paper.

The rest of the paper is organized as follows. We introduce the NMEM model and estimation procedure in Section \ref{ProposedModel}. The real data analysis is reported in Section \ref{RDA}. The simulation and comparisons with other methods are presented in Section \ref{Simulation Study}.

\section{Nonparametric Mixed-Effects Mixture Model} \label{ProposedModel}

\subsection{Model Specification}
Denote $y_{ij}$ as the observation from subject $i$ at time $t_{ij}$ where $i=1,\ldots,m$ and $j=1,\ldots,n_i$. Let $\boldsymbol{y}_i=(y_{i1},\ldots,y_{in_i})^T$ and $\boldsymbol{t}_i=(t_{i1},\ldots,t_{in_i})^T$ be vectors of observations and time points from subject $i$. 
Let $u_{ik}$ be a latent variable such that $u_{ik}=1$ if subject $i$ belongs to group $k$ and $u_{ik}=0$ otherwise. In this paper, for simplicity, we consider two latent groups (e.g., symptomatic and asymptomatic). The extension to more than two groups is straightforward. Denote $\boldsymbol{x}_i$ as a vector of covariates from subject $i$. We assume the following NMEM model:
\begin{align} 
    p_{i1} &:= P(u_{i1}=1)=\frac{\exp(\beta_0+\boldsymbol{x}_i^T \boldsymbol{\beta}_1)}{1+\exp(\beta_0+\boldsymbol{x}_i^T \boldsymbol{\beta}_1)}, \label{MainModel1}\\
    \boldsymbol{y}_{i}&= f_k\left(\boldsymbol{t}_{i}\right)+\mathbf{Z}_{i(k)} \mathbf{b}_{i(k)}+\boldsymbol{\epsilon}_{i} \text{ if } u_{ik}=1, \  k=1,2,  \label{MainModel2}
\end{align}
where $p_{i1}$ is the probability of subject $i$ belonging to group $1$; $f_k(\boldsymbol{t}_i)=(f_k({t}_{i1}),f_k({t}_{i2}),...,\\f_k({t}_{in_1}))^T$ is the mean function in group $k$, $f_k$, evaluated at time points $\boldsymbol{t}_i$; $\mathbf{Z}_{i(k)}$ is the design matrix for random effects; $\mathbf{b}_{i(k)} \stackrel{iid}{\sim} \text{N}(\mathbf{0}, \mathbf{G}_{i(k)})$ are the random effects associated with subject $i$ nested within group $k$; $\boldsymbol{\epsilon}_{i} \sim \text{N}(\mathbf{0},\sigma^2 \mathbf{I}_{n_i})$ are the random errors; and $\mathbf{I}_{n_i}$ is an identity matrix with dimension $n_i$. 
We assume that $u_{ik}$, $\mathbf{b}_{i(k)}$, and $\boldsymbol{\epsilon}_{i}$ from different subjects are independent.

The logistic regression model in equation (\ref{MainModel1}) models the probability of subject $i$ belonging to group $1$ as a function of covariates, and the nonparametric mixed-effect model in equation (\ref{MainModel2}) models longitudinal trajectories of subject $i$ given the group label.  
Given a latent group, since the trajectory of the mean function is usually unknown and could be nonlinear, we model the shape of the mean function nonparametrically using a cubic spline. Specifically, we scale time points into the interval $[0,1]$, denote the first and second derivative of a function $f$ as $f^\prime$ and $f^{\prime \prime}$, and assume that $f_k \in W_2^2[0,1]$ where 
\begin{align}
    W_2^2[0,1] &= \{ f: \ f \text{ and } f^\prime \text{ are absolutely continuous}, \int_{0}^{1} (f^{\prime \prime}(t))^2 dt < \infty \} \nonumber
\end{align}
is a Sobolev space.
The space can be decomposed into two subspaces $W_2^2[0,1] = \mathcal{H}_0 \oplus \mathcal{H}_1$, where $\mathcal{H}_{0}=\left\{f: \ f''=0\right\}$ and $\mathcal{H}_{1}=\left\{f: \ f(0)=0, \ f'(0)=0, \  \int_{0}^{1}\left(f''(t)\right)^{2}<\infty\right\}$ are reproducing kernel Hilbert spaces (RKHS) with reproducing kernels (RK)
$R_{0}(s, t)=1 + k_{1}(s) k_{1}(t)$, 
$R_{1}(s, t)=k_{2}(s) k_{2}(t) - k_{4}(|s-t|)$,
$k_{1}(t)=t-0.5$, $k_{2}(t)=\frac{1}{2}\left(k_{1}^{2}(t)-\frac{1}{12}\right)$, and $k_{4}(t)=\frac{1}{24}\left(k_{1}^{4}(t)-\frac{k_{1}^{2}(t)}{2}+\frac{7}{240}\right)$. See \cite{wang_smoothing_2011} for more details.

The random effects $\mathbf{b}_{i(k)}$ model deviation of subject $i$'s trajectory from the group mean. Different models may be considered for different applications. For example, one may include random intercepts and slopes. Since different subjects may have different nonlinear shapes in our application, we will consider smooth random effects associated with cubic splines \citep{wang_mixed_1998}. Specifically, in addition to random intercepts and slopes, we will consider a zero mean Gaussian process with a covariance function proportional to the RK $R_1(s,t)$. Details are given in Section \ref{RDA}.

\subsection{Model Estimation} \label{ME}

Denote parameters in the covariance matrix of random effects $\mathbf{G}_{i(k)}$ as $\boldsymbol{\zeta}_k$. We need to estimate $\beta_0$, $\boldsymbol{\beta}_1$, $\sigma^2$, $f_1$, $f_2$, $\boldsymbol{\zeta}_1$, and $\boldsymbol{\zeta}_2$. 
Denote $N=\sum_{i=1}^m n_i$ as the total number of observations from all subjects. 
Let $\boldsymbol{y} = (\boldsymbol{y}_1^T,\ldots,\boldsymbol{y}_m^T)^T$, $\boldsymbol{u}_1=(u_{11},\ldots,u_{m1})^T$, and $\mathbf{V}_{ik}=\mathbf{Z}_{i(k)} \mathbf{G}_{i(k)} \mathbf{Z}_{i(k)}^{T} + \sigma^2 \mathbf{I}_{n_i}$. The complete data likelihood of $(\boldsymbol{y}, \boldsymbol{u}_1)$ is calculated as $p(\boldsymbol{y}, \boldsymbol{u}_1) = p(\boldsymbol{y}| \boldsymbol{u}_1) p(\boldsymbol{u}_1) = \prod_{i=1}^{m} p(\boldsymbol{y}_i|u_{i1}) p(u_{i1})$. We estimate all parameters and nonparametric functions using the following penalized likelihood:
{\footnotesize
\begin{align} \label{P-likelihood} 
        l_c 
        =& \left( \sum_{i=1}^{m} \left[ u_{i1} (\beta_0 + \boldsymbol{x}^T_i \boldsymbol{\beta}_1) - \log{(1+\exp{ (\beta_0 + \boldsymbol{x}^T_i \boldsymbol{\beta}_1) })} \right] + \lambda_0 \| \boldsymbol{\beta}_1 \| \right) \nonumber \\
         - & \left(  \sum_{i=1}^{m} u_{i1} \left[ \frac{ n_i}{2} \log{(2 \pi )} + \frac{1}{2} \log{|\mathbf{V}_{i1}| + \frac{1}{2 } \left( \boldsymbol{y}_i -   f_1(\boldsymbol{t}_{i}) \right)^T \mathbf{V}_{i1}^{-1} \left( \boldsymbol{y}_i -   f_1(\boldsymbol{t}_{i}) \right) } \right] + N \lambda_1 \int_{0}^{1}(f_1'')^2 dt  \right) \nonumber \\
         - & \left(  \sum_{i=1}^{m} u_{i2} \left[ \frac{n_i}{2} \log{(2 \pi )} + \frac{1}{2} \log{|\mathbf{V}_{i2}| + \frac{1}{2 } \left( \boldsymbol{y}_i -  f_2(\boldsymbol{t}_{i}) \right)^T \mathbf{V}_{i2}^{-1} \left( \boldsymbol{y}_i -  f_2(\boldsymbol{t}_{i}) \right) } \right] + N \lambda_2 \int_{0}^{1}(f_2'')^2 dt \right),
\end{align}
} where the first part of the first line corresponds to $\log p(\boldsymbol{u}_1)$, the $L_1$ penalty with tuning parameter $\lambda_0$ in the second part of the first line is used to select covariates in $\boldsymbol{x}$, the first parts on the second and third lines correspond to $\log p(\boldsymbol{y}|\boldsymbol{u}_1)$, and the second parts in the second and third lines with smoothing parameters $\lambda_1$ and $\lambda_2$ control the smoothness of the mean functions in two groups.

Since the latent variables $\boldsymbol{u}_1$ are not observed, we use the EM algorithm to estimate the parameters. The E-step of the EM algorithm involves taking the conditional expectation of the likelihood conditional on the data and previously updated parameter values. Denote $\boldsymbol{\theta}=(f_1,f_2,\beta_0,\boldsymbol{\beta}_1,\boldsymbol{\zeta}_1,\boldsymbol{\zeta}_2,\sigma^2,\lambda_0,\lambda_1,\lambda_2)$ as all the parameters to be estimated in both groups. Note that $u_{ik}$ is Bernoulli. The conditional expectation of the latent variable $u_{ik}$ 
\begin{align} \label{Estep}
    w_{i k}
    :=P({u_{ik}}=1|\boldsymbol{y}_i,\hat{\boldsymbol{\theta}})
    =\frac{P({u_{ik}}=1|\hat{\boldsymbol{\theta}}) P(\boldsymbol{y}_i|{u_{ik}}=1,\hat{\boldsymbol{\theta}})}{\sum_{k'=1}^{2} P(u_{ik'}=1|\hat{\boldsymbol{\theta}}) P(\boldsymbol{y}_i|u_{ik'}=1,\hat{\boldsymbol{\theta}})},
\end{align}
where $k \in \{1,2\}$, $\hat{\boldsymbol{\theta}}$ represents the estimated parameters from the last iteration, and $P(\boldsymbol{y}_i|u_{ik}=1,\hat{\boldsymbol{\theta}})$ is a multivariate Gaussian density function of $\boldsymbol{y}_i$ with mean $f_k(\boldsymbol{t}_i)$ and variance $\mathbf{V}_{ik}$.  

The conditional expectation of $l_c$ becomes 
{\footnotesize
\begin{align} 
& \mathbb{E} (l_c \mid \boldsymbol{y}, \hat{\boldsymbol{\theta}}) = \nonumber\\
\begin{split}
    & \left( \sum_{i=1}^{m} \left[ w_{i1} (\beta_0 + \boldsymbol{x}^T_i \boldsymbol{\beta}_1) - \log{(1+\exp{ (\beta_0 + \boldsymbol{x}^T_i \boldsymbol{\beta}_1) })} \right] + \lambda_0 \| \boldsymbol{\beta}_1 \| \right) \label{logistic} 
\end{split}
\\
\begin{split}  
    - &  \left(  \sum_{i=1}^{m} w_{i1} \left[ \frac{ n_i}{2} \log{(2 \pi )} + \frac{1}{2} \log{|\mathbf{V}_{i1}| + \frac{1}{2 } \left( \boldsymbol{y}_i -   f_1(\boldsymbol{t}_{i}) \right)^T \mathbf{V}_{i1}^{-1} \left( \boldsymbol{y}_i -   f_1(\boldsymbol{t}_{i}) \right) } \right] + N \lambda_1 \int_{0}^{1}(f_1'')^2 dt  \right) \nonumber
\end{split}
\\
\begin{split}
    - & \left(  \sum_{i=1}^{m} w_{i2} \left[ \frac{ n_i}{2} \log{(2 \pi )} + \frac{1}{2} \log{|\mathbf{V}_{i2}| + \frac{1}{2 } \left( \boldsymbol{y}_i - f_2(\boldsymbol{t}_{i}) \right)^T \mathbf{V}_{i2}^{-1} \left( \boldsymbol{y}_i - f_2(\boldsymbol{t}_{i}) \right) } \right] + N \lambda_2 \int_{0}^{1}(f_2'')^2 dt \right) \label{VARAndCD}
\end{split}
\end{align}
}
where $w_{i2} = 1 - w_{i1}$ and  $\hat{\boldsymbol{\theta}}$ represents the estimated parameters from the last iteration.

The M-step involves maximizing $E(l_c \mid \boldsymbol{y}, \hat{\boldsymbol{\theta}})$. We maximize equations (\ref{logistic}) and (\ref{VARAndCD}) separately since they involve two disjoint set of parameters
$\boldsymbol{\theta}_1=(\beta_0,\boldsymbol{\beta}_1,\lambda_0)$ and 
$\boldsymbol{\theta}_2=(f_1, f_2,\boldsymbol{\zeta}_1,\boldsymbol{\zeta}_2,\sigma^2,\lambda_1,\lambda_2)$ respectively. 

Equation (\ref{logistic}) is the penalized likelihood of a logistic regression model with an $L_1$ penalty. We apply the existing method and software package in R \citep{friedman_regularization_2010} to update the estimate of $\boldsymbol{\theta}_1$. We use 10-fold cross-validation to select the tuning parameter $\lambda_0$. 

For equation (\ref{VARAndCD}), the mean functions $f_1$ and $f_2$ are modeled using cubic splines. Based on the decomposition of the Sobolev space $W_{2}^{2}[0,1] = \mathcal{H}_0 \oplus \mathcal{H}_1$, the estimated mean functions can be expressed as \citep{wang_smoothing_2011}
\begin{align}
\label{decompose_mean}
    f_k(t_{ij})=\sum_{\nu=1}^2 d_{k\nu} \phi_{\nu}(t_{ij})+\sum_{l=1}^e  c_{kl} \xi_l(t_{ij}), \quad  k=1,2,
\end{align}
where $\phi_1(t)=1$ and $\phi_2(t)=t$ are basis functions of $\mathcal{H}_0$, $\xi_l(t)=R_1(z_l,t)$, and $\{z_1, \cdots,z_e\}$ are $e$ distinct points in  the set $\{t_{ij}, \ i=1,\ldots,m, \ j=1,\ldots,n_i\}$. 
Let $\boldsymbol{d}_{k}=(d_{k1},d_{k2})^T$, 
$\boldsymbol{c}_{k}=(c_{k1},\ldots,c_{ke})^T$, and $(\tau_1,\ldots,\tau_N)^T$ be the stacked vectors of all time points $(t_{11},\ldots,t_{1n_1},\ldots,\\t_{mn_m})^T$. Denote $\mathbf{S}$ as an $N \times 2$ matrix with the $(\iota,\nu)$th entry $\phi_\nu(\tau_\iota)$, $\mathbf{R}$ as an $N \times e$ matrix with $(\iota,l)$th entry $R_1(\tau_\iota,z_l)$, and $\mathbf{Q}$ as an $e \times e$ matrix with the $(l,k)$th entry $R_1(z_l,z_k)$. Let $\mathbf{W}_k$ be a block diagonal matrix of size $N \times N$ where the $i$-th block is $w_{ik} \mathbf{V}_{ik}^{-1}$. It is easy to verify that the target equation (\ref{VARAndCD}) is proportional to
\begin{align} 
& \sum_{i=1}^{m}\left( w_{i 1} \log \left|\mathbf{V}_{i 1}\right| \right) + \left(\boldsymbol{y}-\mathbf{S} \boldsymbol{d}_{1}-\mathbf{R} \boldsymbol{c}_{1}\right)^{T} \mathbf{W}_{1}\left(\boldsymbol{y}-\mathbf{S} \boldsymbol{d}_{1}-\mathbf{R} \boldsymbol{c}_{1}\right)+N \lambda_{1} \boldsymbol{c}_{1}^{T} \mathbf{Q} \boldsymbol{c}_{1} \label{f1} \\
+ & \sum_{i=1}^{m}\left( w_{i 2} \log \left|\mathbf{V}_{i 2}\right| \right)+ \left(\boldsymbol{y}-\mathbf{S} \boldsymbol{d}_{2}-\mathbf{R} \boldsymbol{c}_{2}\right)^{T} \mathbf{W}_{2}\left(\boldsymbol{y}-\mathbf{S} \boldsymbol{d}_{2}-\mathbf{R} \boldsymbol{c}_{2}\right)+N \lambda_{2} \boldsymbol{c}_{2}^{T} \mathbf{Q} \boldsymbol{c}_{2}. \label{f2}
\end{align}
We estimate the variance components $(\sigma^2,\boldsymbol{\zeta}_1, \boldsymbol{\zeta}_2)$ and components corresponding to the mean functions $(\boldsymbol{c}_{1}, \boldsymbol{d}_{1}, \lambda_{1}, \boldsymbol{c}_{2}, \boldsymbol{d}_{2}, \lambda_{2})$
alternatively.

When fixing the variance components $(\sigma^2,\boldsymbol{\zeta}_1, \boldsymbol{\zeta}_2)$, we estimate the two mean functions and their smoothing parameters in equation (\ref{f1}) and (\ref{f2}) separately. Each one is the penalized least square for smoothing spline regression with correlated data. The minimizer of each satisfies the following equations \citep{gu_smoothing_2013}:
\begin{align} \label{weightedSystem}
    \left(\begin{array}{cc}
    \mathbf{S}^{T} \mathbf{W}_k \mathbf{S} & \mathbf{S}^{T} \mathbf{W}_k \mathbf{R} \\
    \mathbf{R}^{T} \mathbf{W}_k \mathbf{S} & \mathbf{R}^{T} \mathbf{W}_k \mathbf{R}+ N \lambda_k \mathbf{Q}
    \end{array}\right)\left(\begin{array}{l}
    \boldsymbol{d}_k \\
    \boldsymbol{c}_k
    \end{array}\right)=\left(\begin{array}{c}
    \mathbf{S}^{T} \mathbf{W}_k \boldsymbol{y} \\
    \mathbf{R}^{T} \mathbf{W}_k \boldsymbol{y}
    \end{array}\right), \quad k=1, 2.
\end{align}
Note that $\mathbf{W}_k$ is fixed at this step.  
Let $w_{ik}\mathbf{V}_{ik}^{-1}=\mathbf{P}_{ik}^T \mathbf{P}_{ik}$ be 
the Cholesky decomposition of $w_{ik}\mathbf{V}_{1k}^{-1}$ and 
$\mathbf{P}_k=\text{diag} (\mathbf{P}_{1k},\cdots,\mathbf{P}_{mk})$.
Then $\mathbf{W}_k=\mathbf{P}_k^{T} \mathbf{P}_k$ is the Cholesky decomposition of 
$\mathbf{W}_k=\text{diag}(w_{1k} \mathbf{V}_{1k}^{-1},\cdots,w_{mk} \mathbf{V}_{mk}^{-1})$.  
With the transformations $\left(\boldsymbol{y}_{k}, \mathbf{S}_{k}, \mathbf{R}_{k}\right)=\mathbf{P}_k^{T}(\boldsymbol{y}, \mathbf{S}, \mathbf{R})$, the weight matrices $\mathbf{W}_k$ in equation (\ref{weightedSystem}) are absorbed into other vectors/matrices and the equations reduce to those under the independent cases. Therefore, we can apply computational methods in \cite{gu_smoothing_2013} for independent data to update $\boldsymbol{c}_k$ and $\boldsymbol{d}_k$ with smoothing parameters $\lambda_k$ selected by the generalized maximum likelihood method. More recent advances in computational methods can be found in \cite{ma_efficient_2015} and \cite{sun_asymptotic_2021}.

When fixing the mean functions, we estimate the variance components by minimizing (\ref{f1}) and (\ref{f2}) together using the Limited-memory Broyden–Fletcher–Goldfarb–Shanno (L-BFGS-B) algorithm \citep{byrd_limited_1995,zhu_algorithm_1997}. Since both equations contain the common variance of random error $\sigma^2$, we profiled it out and minimized the profiled likelihood. We provide the calculation of profiled likelihood in the appendix.

The initial estimates of variance components and the mean function are calculated using the linear mixed effects (LME) form of smoothing splines. These estimates are then used for the first inner iteration. The existing package nlme can be used for fitting these LME models. Details can be found in \cite{wang_smoothing_1998} and \cite{xu_lowrank_2021}.

We summarize the entire EM algorithm in Algorithm \ref{Algo1}. The stopping criteria are set as follows. 
Let $\boldsymbol{\tilde{\theta}}$ represent all the estimated values except the penalty parameter $\lambda_0$ and the smoothing parameters $\lambda_1$ and $\lambda_2$. Let $d_{EM}^{[\mathcal{O}]} =  \frac{\|\boldsymbol{\tilde{\theta}}^{[\mathcal{O}]}-\boldsymbol{\tilde{\theta}}^{[\mathcal{O}-1]} \|^2}{\| \boldsymbol{\tilde{\theta}}^{[\mathcal{O}-1]}  \|^2 + \kappa_1} $ be the relative change of the estimates at the ${\mathcal{O}}$th EM iteration and $d_{inner}^{[\mathcal{O},\mathcal{I}]} =  \frac{\|\boldsymbol{\tilde{\theta}}^{[\mathcal{O},\mathcal{I}]}-\boldsymbol{\tilde{\theta}}^{[\mathcal{O}, \mathcal{I}-1]} \|^2}{\| \boldsymbol{\tilde{\theta}}^{[\mathcal{O}, \mathcal{I}-1]}  \|^2 + \kappa_2} $ be the relative change of the estimates at the $\mathcal{I}$th inner iteration inside the ${\mathcal{O}}$th EM iteration. The inner iteration stops if $d_{inner}^{[\mathcal{O},\mathcal{I}]} \leq D_{inner}$ and the EM iteration stops if $d_{EM}^{[\mathcal{O}]} \leq D_{EM}$. The maximum number of iterations are denoted as $\mathcal{O}_{max}$ and $\mathcal{I}_{max}$, respectively. 

\IncMargin{1em}
\begin{algorithm} 
\label{Algo1}
\SetAlgoLined
\SetKwInOut{Input}{Input}\SetKwInOut{Output}{Output}
\Indm
\Input{longitudinal data}
\Output{estimates of group probabilities, variance components, parameters of covariates and mean functions: $\hat{p}_{ik}$, $\hat{\boldsymbol{\zeta}}_k$, $\hat{\sigma}^2$, $\hat{\beta_0}$, $\hat{\boldsymbol{\beta}_1}$,  $\hat{f}_k$}
\Indp
randomly assign subjects with probability 0.5 into two groups where group 1 contains $n_1$ subjects and group 2 contains $n_2$ subjects\;
set $\hat{p}_{ik} = n_k/N$ for all $i$\;
get the initial estimates of $\hat{\boldsymbol{\zeta}}_k,\hat{\sigma}^2$ and $\hat{f}_k$ using the method introduced in \cite{wang_smoothing_1998} and \cite{xu_lowrank_2021}\;
\While{$(\{d_{EM}^{[\mathcal{O}]} > D_{EM}\} \wedge \{\mathcal{O} <= \mathcal{O}_{max}\})$}{
    \textbf{E-step:} \\
    compute $w_{ik}$ as in equation (\ref{Estep}) using the current estimates $\hat{p}_{ik}$, $\hat{\boldsymbol{\zeta}}_k$, $\hat{\sigma}^2$ and $\hat{f}_k$\;
    \textbf{M-step:} \\
    update $\hat{\boldsymbol{\beta}}$ using logistic regression as in equation (\ref{logistic}) and calculate $\hat{p}_{ik}$ as in equation (\ref{MainModel1})\; 
    \While{$(\{d_{inner}^{[\mathcal{O},\mathcal{I}]} > D_{inner}\} \wedge \{\mathcal{I} <= \mathcal{I}_{max}\})$}{
        solve equation (\ref{f1}) and (\ref{f2}) separately to update $f_k$ with fixed $\hat{\boldsymbol{\zeta}}_k$ and $\hat{\sigma}^2$\;
        minimize the sum of equation (\ref{f1}) and (\ref{f2}) to update $\hat{\boldsymbol{\zeta}}_k$ and $\hat{\sigma}^2$ with fixed $f_k$ using L-BFGS-B\;
    }
}
\Return $\hat{p}_{ik}$, $\hat{\boldsymbol{\zeta}}_k$, $\hat{\sigma}^2$, $\hat{\beta_0}$, $\hat{\boldsymbol{\beta}_1}$,  $\hat{f}_k$
\caption{EM algorithm for fitting the NMEM model}
\end{algorithm}
\DecMargin{1em}

We regard the $L_1$ penalty in the logistic regression as a variable selection process. 
After variable selection, we rerun Algorithm \ref{Algo1} without the $L_1$ penalty and get the point estimate of all parameters. 
We construct bootstrap confidence intervals for the estimated mean functions and all parameters. 
We note that even though we are only interested in two clusters for our specific problem, the number of clusters in other applications is usually unknown. Model selection methods, such as BIC described in \cite{ma_penalized_2008}, can be used to select the number of clusters.

\section{Temperature Profiles in COVID-19-Infected HD Patients} \label{RDA}

This section applies our methods to investigate temperature change profiles before COVID-19 confirmation in HD patients. We consider 3,293 ESRD patients who received in-center HD treatment from Fresenius Medical Care and had positive PCR tests during 2020-01-01 and 2021-8-31. We align patients at their first positive PCR test date and analyze their temperature measurements 30 days before the PCR test. The observation window is defined as -30 to 0, where 0 is the PCR test date. The average number of observations for each patient is around 14. To focus on the changing pattern, we subtract each patient's temperature from the average temperature between -60 to -31 days, a period free of COVID-19 infection. Figure \ref{figPre} shows temperature change profiles for all patients. Our goals are to identify two latent groups (symptomatic and asymptomatic), estimate the probability of belonging to each group for each subject, and associate the group probability with demographic and clinical characteristics.  

Based on preliminary analyses, we consider an NMEM model with equation (\ref{MainModel1}) and the following model in place of equation (\ref{MainModel2}):

\begin{align}
\label{tempmodel}
    \boldsymbol{y}_i &= f_{k}\left(\boldsymbol{t}_{i}\right)+b_{i 1(k)}  \mathbf{1}+b_{i 2(k)}  \boldsymbol{t}_{i}+s_{i(k)}\left(\boldsymbol{t}_{i}\right)+\boldsymbol{\epsilon}_{i}, \text{ if $u_{ik}=1$, } k=1,2, 
\end{align}
where groups $1$ and $2$ correspond to patients with and without temperature changes, $f_1$ and $f_2$ are the mean functions of these two groups, $b_{i 1(k)}$, $b_{i 2(k)}$, and $s_{i(k)}(\boldsymbol{t}_i)=(s_{i(k)}(t_{i,1}), \cdots, s_{i(k)}(t_{i,n_i}))^T$ are random intercept, slope, and a vector of smooth random effects associated with subject $i$ nested within group $k$, and 
$\boldsymbol{\epsilon}_{i} \sim \text{N}(\mathbf{0}, \sigma^2 \boldsymbol{I}_{n_i})$ are random errors independent of the random effects. Model \eqref{tempmodel} is a special case of the proposed model \eqref{MainModel2} with 
$\mathbf{Z}_{i(k)}=(\boldsymbol{1}, \boldsymbol{t}_i, \mathbf{I}_{n_i})$
and 
$\boldsymbol{b}_{i(k)}=(b_{i1(k)}, b_{i2(k)}, s_{i(k)}(t_{i,1}), \cdots, s_{i(k)}(t_{i,n_i}))^T$.
We assume unstructured covariance for the random intercept and slope.
In addition to the random intercept and slope, the smooth random effect $s_{i(k)}(t)$ is a function of $t$ that allows a more flexible deviation of subject $i$ from the population mean \citep{wang_smoothing_2011}. We assume that $s_{i(k)}(t)$ follows a zero mean Gaussian process with a covariance function equal to the reproducing kernel of $\mathcal{H}_1$ \citep{wang_mixed_1998}. Specifically, for the combined random effects, we assume that     

\begin{align}
    \boldsymbol{b}_{i(k)} &= 
    \left(\begin{array}{c}
    b_{i1(k)} \\
    b_{i2(k)} \\ \hline
    s_{i(k)}(t_{i,1}) \\
    \vdots \\
    s_{i(k)}(t_{i,n_i})
    \end{array} \right)
    \sim \text{N} 
    \left ( 
    \boldsymbol{0},
    \left[
    \begin{array}{c|c}
    \begin{array}{cc}
    \sigma^2_{inter,k} & \sigma_{is,k} \\
    \sigma_{is,k} & \sigma^2_{slope,k} \\
    \end{array} & \mathbf{0} \\ \hline
    \mathbf{0} & \begin{array}{cc}
    \sigma^2_{non,k} R_{1}(\boldsymbol{t}_i,\boldsymbol{t}_i)
    \end{array}
    \end{array}
    \right]
     \right ), \label{RE1}
\end{align}
for $k \in \{1,2\}$, where $\sigma^2_{inter,k}$, $\sigma^2_{slope,k}$, and $\sigma_{is,k}$ are variances and covariance of the random intercept and slope, $\sigma^2_{non,k}$ is the variance of smooth random effect, and $R_{1}(\boldsymbol{t}_i,\boldsymbol{t}_i)$ is an $n_i \times n_i$ matrix with the $jk$th entry as $R_1(t_{ij},t_{ik})$. For model \eqref{tempmodel}, we have $\boldsymbol{\zeta}_k = (\sigma^2_{inter,k}, \sigma_{is,k}, \sigma^2_{slope,k}, \sigma^2_{non,k})$, $k \in \{1,2\}$.

Based on exploratory analysis and literature, we consider the following covariates for the logistic regression model (\ref{MainModel1}): 
gender, race, ethnicity, vintage, diabetes, hypertension, BMI (calculated using height and the average weight after each dialysis treatment during the observation window), age on the treatment date, and vascular access type with three options: arteriovenous fistula (AVF), arteriovenous grafts (AVG), and central venous catheter (CVCATH).

For the stopping criteria, we set $\kappa_1 = \kappa_2 = 10^{-5}$, $D_{EM} = 10^{-5}$, $D_{inner} = 10^{-5}$, $\mathcal{O}_{max}=100$, and $\mathcal{I}_{max}=5$. 
Different hyperparameters parameters lead to similar fits. 
After variable selection, we refit the model with selected variables and without the $L_1$ penalty. To avoid over-fitting, we set a lower bound for $\lambda_k$ as $\log_{10} (N\lambda_k) = 0$ where $N=46837$. The variance components in equation (\ref{RE1}) are estimated using the covariance matrix of bivariate normal where $\sigma_{is,1} = \rho \sigma_{inter,1} \sigma_{slope,1}$. All variances are estimated using natural log transformation, and a tangent transformation is used for the correlation parameter $\rho$.

We perform the following diagnostics to evaluate the model assumptions. First, to check the independence assumption between random errors in $\boldsymbol{\epsilon}_{i}=(\epsilon_{i1},\cdots,\epsilon_{in_i})^T$ and model potential temporal correlation within patient $i$, we consider a continuous AR(1) random error structure with 
covariance between $\epsilon_{ik}$ at time $t_{ik}$ and $\epsilon_{il}$ at time $t_{il}$ equals $\sigma^2 \phi^{|t_{ik} - t_{il}|}$. We extend our estimation and computational methods to fit models with correlated random errors. The estimate of $\phi$ equals $2.85 \times 10^{-297}$ with 95\% bootstrap confidence interval [$6.02*10^{-307}$,$3.86*10^{-38}$], indicating the autocorrelation is ignorable. Second, to check the normality assumptions for random errors and random effects, we compute estimates of random effects using the following formula in \cite{wang_mixed_1998}:
\begin{align*}
    \mathbf{Z}_{i(k)} \hat{\mathbf{b}}_{i(k)}=\mathbf{Z}_{i(k)} \hat{\mathbf{G}}_{i(k)} \mathbf{Z}_{i(k)}^{\prime} \hat{\mathbf{V}}_{ik}^{-1} (\boldsymbol{y}-\mathbf{S} \hat{\boldsymbol{d}}_k- \mathbf{R} \hat{\boldsymbol{c}}_k) ,
\end{align*}
where $\hat{\mathbf{G}}_{i(k)}$,$\hat{\mathbf{V}}_{ik}$, $\hat{\boldsymbol{d}}_k$, and $\hat{\boldsymbol{c}}_k$ are the estimates, and a threshold of $0.5$ was used to cluster patients into two groups. The estimated random effects and residuals' QQ plots (not shown) show no evidence of substantial violations of the normality assumptions.

We summarize estimates of coefficients associated with covariates, variance components, and their 95\% bootstrap confidence intervals based on $1000$ samples from the fitted model in Table \ref{Tab:RD}. 

We conclude that white and elderly patients were less likely to have a change in temperature. Age can reflect the strength of the patient's immune system. Elderly patients usually have a weaker immune system and are thus less responsive to the infection.

\begin{table}
\caption{Result of data analysis. Upper: estimates of variance components. Lower: estimates of coefficients in the logistic regression model. 
We only report the variables selected by Lasso. We report the point estimate and 95\% bootstrap confidence intervals. The subscripts for variance components are defined in equation (\ref{RE1}). The subscripts for covariates with a ``Y" indicate the patient has that comorbidity. }
\centering
\begin{tabular}{ V{3} c|r|c V{3} } 
\specialrule{.2em}{.1em}{.1em}
Parameters & Estimate & 95\% CI \\
\hline 
$\hat{\sigma}^2$&0.4136&(0.4027,0.4239)\\
\hline
$\hat{\sigma}_{inter,1}^2$&0.0955&(0.0571,0.1854)\\
\hline
$\hat{\sigma}_{inter,2}^2$&0.0620&(0.0000,0.0752)\\
\hline
$\hat{\sigma}_{is,1}$&-0.0809&(-0.1435,-0.0291)\\
\hline
$\hat{\sigma}_{is,2}$&-0.0135&(-0.0268,0.0053)\\
\hline
$\hat{\sigma}_{slope,1}$&0.5130&(0.3793,0.6539)\\
\hline
$\hat{\sigma}_{slope,2}$&0.0956&(0.0467,0.1251)\\
\hline
$\hat{\sigma}_{non,1}^2$&14.8065&(9.0019,20.3593)\\
\hline
$\hat{\sigma}_{non,2}^2$&0.2195&(0.0000,1.0791)\\
\specialrule{.1em}{.05em}{.05em}
$\hat{\beta}_0$&-1.0028&(-1.6899,0.2948)\\
\hline
$\hat{\beta}_{White}$&-0.3734&(-0.625,-0.0432)\\
\hline
$\hat{\beta}_{DiabetesY}$&-0.1085&(-0.3421,0.1256)\\
\hline
$\hat{\beta}_{BMI}$&0.0139&(-0.0047,0.0268)\\
\hline
$\hat{\beta}_{age}$&-0.0107&(-0.0203,-0.0023)\\
\hline
$\hat{\beta}_{AVG}$&0.2057&(-0.1251,0.5811)\\
\hline
$\hat{\beta}_{CVCATH}$&-0.1675&(-0.5195,0.1863)\\
\specialrule{.2em}{.1em}{.1em}
\end{tabular}
\label{Tab:RD}
\end{table}

\begin{figure}
    \centering
    \includegraphics[scale=0.4]{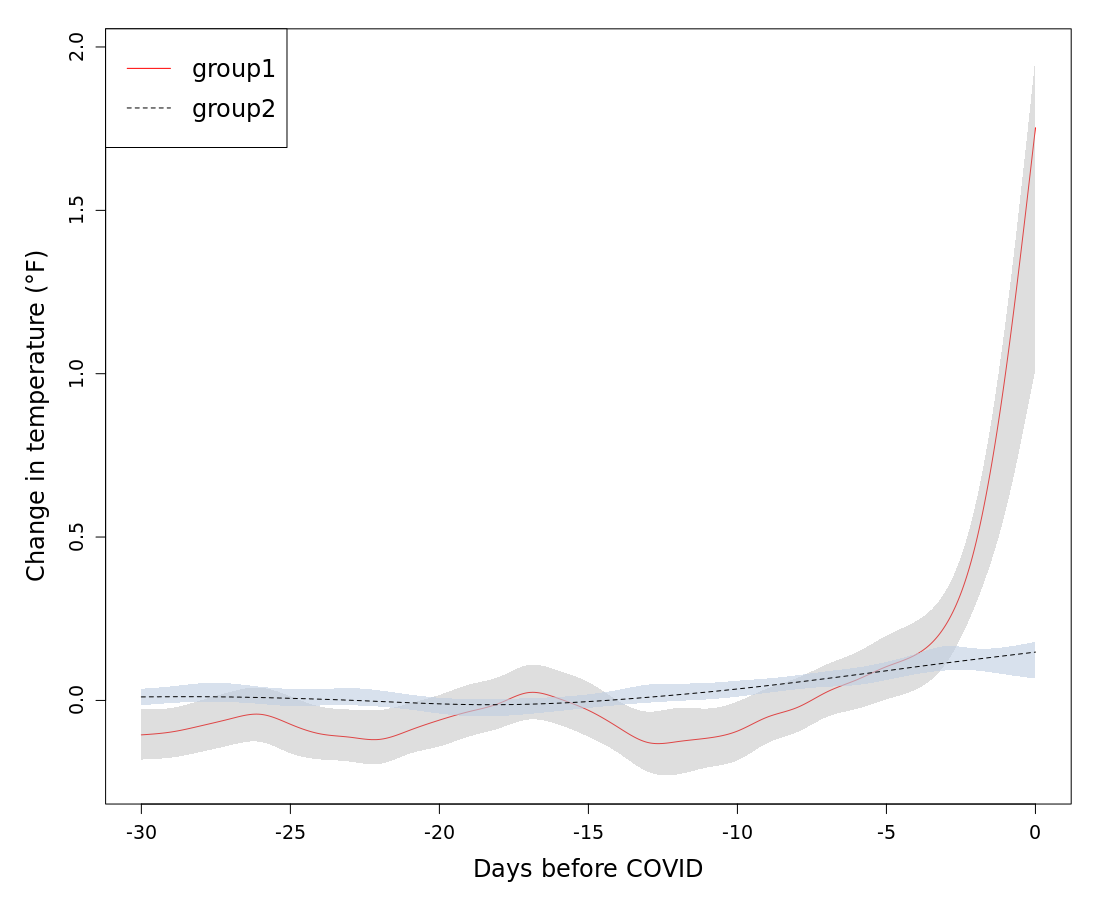}
    \caption{Estimated mean functions of two groups and their 95\% bootstrap confidence intervals.}
    \label{figRD1}
\end{figure}

With the probability threshold for clustering set as $0.5$, our method identified 468 (14.21\%) out of 3,293 COVID-19 patients who experienced an increase in temperature before the positive PCR test. Figure \ref{figRD1} shows the estimated mean change functions and their 95\% confidence intervals in the two groups. We note that confidence intervals for the mean functions are narrow due to a large number of total observations. As we can see, in group 1, the increase in temperature started about nine days and accelerated about four days before the PCR test date.

Table \ref{Tab:profile} shows several interesting characteristics that set patients with fever apart from those who did not have fever. Two observations are striking: first, age is higher in patients who did not have fever, and second, the rate of whites without fever was higher compared to the ones with fever. The observation regarding age agrees with clinical experience. Fever is less prevalent in the elderly ($>65$ years) than in younger adults. The body temperature in the elderly is physiologically lower compared to middle-aged adults.
In frail, older adults, fever is absent in 30 to 50 percent, even in serious infections such as pneumonia or endocarditis \citep{Musgrave_clinical_1990,henschke_infections_1993}. The blunted febrile response in older adults is mostly due to impairment in multiple systems responsible for thermoregulation (e.g., shivering, vasoconstriction, hypothalamic regulation, and thermogenesis by brown adipose tissue) \citep{Mackowiak_1997}. 
The impact of race on fever is less clear. 
Interestingly, in a study, fever was less prevalent in blacks compared to whites with temporal measurement and more prevalent with oral measurement
\citep{bhavani_racial_2022}. In that study of patients with suspected (non-COVID) infection, in 265 black patients with paired measurements within the first hour of presentation, fever prevalence was 23.4\% with temporal and 35.8\% with oral measurement. In 281 white patients, the prevalence was 27.8\% with temporal and 26.0\% with oral measurement.

\begin{table}
\caption{Statistics of patients in different groups. The abbreviation "std" stands for standard deviation. }
\centering
\begin{tabular}{ V{3} c|c|c|c V{3} } 
\specialrule{.2em}{.1em}{.1em}
Variable & All & Temperature Increased & Temperature Unchanged \\
\hline
Female (\%)&	44&	42	&44\\
\hline
White (\%)	&64	&55&	66\\
\hline
Hispanic (\%)&	21&	20&	21\\
\hline
Diabetes (\%)&	52&	49&	52\\
\hline
Hypertension (\%)	&79&	80&	79\\
\hline
AVF (\%)&	67	&68	&67\\
\hline
AVG (\%)&	13	&15&	12\\
\hline
BMI (std)	&30.84 (8.20)&	31.78 (9.35)	&30.68 (7.99)\\
\hline
Age (std)	&61.69 (14.19)	&59.58 (14.03)&	62.04 (14.19)\\
\hline
Vintage (std) &	4.64 (4.09)&	4.69 (4.40)&	4.64 (4.04)\\
\specialrule{.2em}{.1em}{.1em}
\end{tabular}
\label{Tab:profile}
\end{table}

\section{Simulation Study} \label{Simulation Study}

We conduct simulations to evaluate the performance of the proposed method and compare it with the previous work in \cite{lu_finite_2012}. \cite{lu_finite_2012} considered the following model:   
\begin{align}
    P(u_{ik}=1) = p_{ik} = 
     \frac{\exp (\beta_0 + \boldsymbol{x}_{i}^{T} \boldsymbol{\beta}_{k})}{\sum_{k'=1}^{2} \exp (\beta_0 +\boldsymbol{x}_{i}^{T} \boldsymbol{\beta}_{k'})}, \quad i = 1,\ldots,m, \nonumber \\
     \boldsymbol{y}_{i}=\tilde{\mathbf{X}}^{[1]}_i f_k(\mathbf{t}_i) + \tilde{\mathbf{X}}^{[2]}_i \boldsymbol{\alpha}_k + \mathbf{Z}_{i(k)} \mathbf{b}_{i(k)} + \boldsymbol{\epsilon}_{ik} \quad \text { if } u_{ik}=1 , 
     \label{LSmodel}
\end{align}
where $\tilde{\mathbf{X}}^{[1]}_i$ and $\tilde{\mathbf{X}}^{[2]}_i$ are the design matrices of fixed effects $f_k(\mathbf{t}_i)$ and $\boldsymbol{\alpha}_k$ respectively, $\mathbf{Z}_{i(k)}$ is the design matrix for the random effects $\mathbf{b}_{i(k)}$, and the variances of random errors $\boldsymbol{\epsilon}_{ik}$ are assumed to be different for different groups.

Our method differs from Lu and Song's methods in both the model structure and estimation approach. \cite{lu_finite_2012} modeled the nonparametric functions $f_k$'s  using P-splines while we use  smoothing splines. Our proposed model includes a smooth random effect for flexibility and allows different random effects in different groups. \cite{lu_finite_2012} estimate parameters in a Bayesian framework while we estimate parameters using penalized likelihood. In addition, our estimation procedure includes an $L_1$ penalty for variable selection.

We generate data using model \eqref{MainModel1} and \eqref{tempmodel}, which was used in real data analysis in Section \ref{RDA}. We set $m=3293$, $n_i$'s are randomly selected from integers in the interval $[16, 31]$, and parameters as their estimates when possible. 

We generate latent variables $u_{i1}$ for $i=1,\cdots,m$ using equation \eqref{MainModel1}. 
We include all nine covariates (before variable selection) for the group probability in equation (\ref{MainModel1}) and generated them according to their estimated marginal distributions. Specifically, six categorical variables are generated according to their empirical ratios in each category. Three continuous variables are generated from an exponential distribution with a rate parameter of $0.22$ (vintage), a Gamma distribution with a shape parameter of $15.58$ and a rate parameter of $0.51$ (BMI), and a normal distribution with a mean $61.69$ and standard deviation $14.20$ (age). Parameters in three continuous distributions are set to be the maximum likelihood estimates. 
For the $\boldsymbol{\beta}$ coefficients in the logistic model (\ref{MainModel1}), we set the intercept as the estimate in Table \ref{Tab:RD}, coefficients for gender, ethnicity, vintage, diabetes, hypertension, BMI, and vascular access types as zero since they are either not selected in the variable selection process or the confidence interval contains zero in the final estimation. The coefficients for race and age are set to be the estimates in Table \ref{Tab:RD}. The latent variables $u_{i1}$ for $i=1,\cdots,m$ are generated according to a Bernoulli distribution with probability given in equation (\ref{MainModel1}). 

We use equation (\ref{tempmodel}) to generate responses $\boldsymbol{y}_i$. 
Since \cite{lu_finite_2012}'s model does not have a smooth random effect, we consider two simulation settings:
with smooth random effect where $\sigma_{non,1}^2=14.8$ (estimate from the real data), $\sigma_{non,2}^2=14$ and without smooth random effect where $\sigma_{non,1}^2=\sigma_{non,2}^2=0$.
In both settings, the values of all other parameters apart from the variance of smooth random effect are set to be the estimates in the real data analysis in Table \ref{Tab:RD}. 
We first randomly generate the number of observations $n_i$ from a discrete uniform distribution on $[16,31]$ for each subject $i$, and then randomly select $n_i$ days between -30 to 0 as $\boldsymbol{t}_{i}$. The values of mean functions $f_k$ at each time point are set to be the estimates in the real data analysis. Independent and identically distributed random errors are generated from a normal distribution with mean zero and variance equals the estimate from the real data analysis.

To fit model \eqref{LSmodel}, we set $\tilde{\mathbf{X}}^{[1]}_i$ as an identity matrix and $\tilde{\mathbf{X}}^{[2]}_i = 0$. \cite{lu_finite_2012} used the random permutation sampler approach in \cite{fruhwirth-schnatter_markov_2001} to deal with the label-switching problem caused by the symmetric prior of the parameters in different components. We use the variance of random slope for the permutation sampler in 
our simulations.

\begin{figure} 
    \centering
    \includegraphics[scale=0.4]{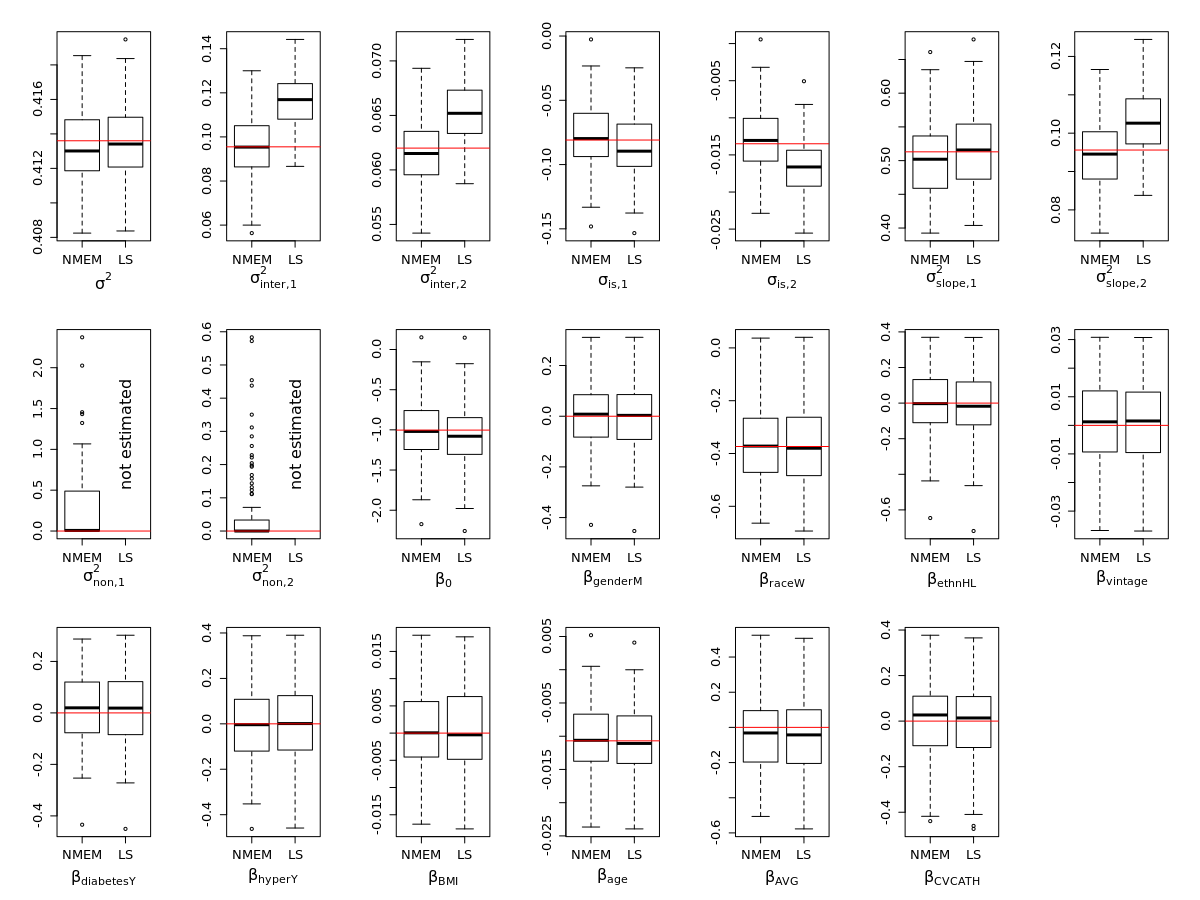}
    \caption{Simulation results without smooth random effects: box plots of estimated parameters from NMEM and Lu and Song's method (LS). Red lines are true values of parameters.} 
    \label{box12noNP}
\end{figure}

\begin{figure} 
    \centering
    \includegraphics[scale=0.4]{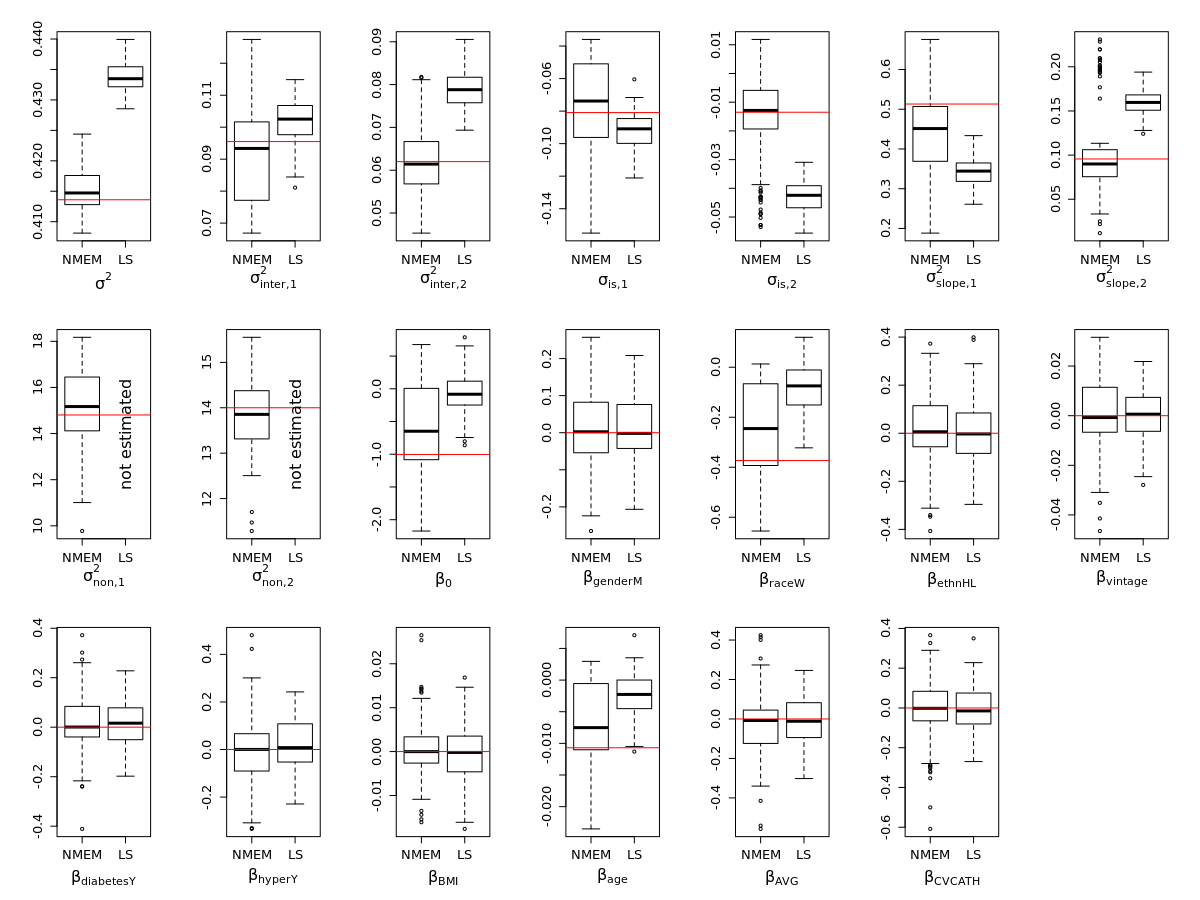}
    \caption{Simulation results with smooth random effects: box plots of estimated parameters from NMEM and Lu and Song's method (LS). Red lines are true values of parameters.} 
    \label{box12largeNP14}
\end{figure}

Each simulation is replicated $100$ times. We use the same stopping criteria as in the real data analysis. The comparison results are presented in Table \ref{Tab:ACC}, Figure \ref{box12noNP} and Figure \ref{box12largeNP14}. 

Both methods performed well when no smooth random effect existed. The clustering accuracy and MSEs of function estimates are almost identical.  The existence of a smooth random effect allows nonlinear individual departure from the population mean function, thus making clustering and estimation more difficult. This is reflected in the smaller accuracy and larger MSEs in Table \ref{Tab:ACC}. As expected, our method achieves better accuracy and lower MSEs than Lu and Song's method when smooth random effect existed.
In general, our method has smaller biases in the estimates of variance components (Figures \ref{box12noNP} and \ref{box12largeNP14}).

\begin{table}
  \centering
  \caption{Simulation results: average accuracy in the two settings. NMEM stands for our method, LS stands for LU and Song's method. We also report the average mean squared error (MSE) of estimated mean functions.}
    \label{Tab:ACC}
  \begin{tabular}{V{3} c|c|c|c|c V{3}}
    \hlineB{3}
            & \multicolumn{2}{c|}{\textbf{Without $\sigma^{2}_{non,k}$}} & \multicolumn{2}{c V{3}}{\textbf{With $\sigma^{2}_{non,k}$}} \\
    \cline{2-5} 
            & NMEM & LS & NMEM & LS \\
    \hlineB{2}
    Accuracy & 0.9341 &  \textbf{0.9344} & \textbf{0.8239} & 0.6351 \\ 
    MSE $f_1$ & \textbf{0.0017} & 0.0018 & \textbf{0.0240} & 0.0494 \\ 
    MSE $f_2$ & \textbf{0.0000} & 0.0001 & \textbf{0.0007} & 0.0158 \\ 
    \hlineB{3}
    \end{tabular}
\end{table}

We also conducted a simulation study to test the performance of variable selection. The values of all parameters, including the variance of the smooth random effects, are set to be the estimates in the real data analysis in table \ref{Tab:RD}. For variable selection in the model \eqref{MainModel1}, on average,  $0.3$ of the two non-zero parameters (not including the intercept) are mistakenly excluded from the model, and $2.95$ out of the eight zero parameters are mistakenly selected.
The over-selection behavior agrees with the previous literature \citep{tibshirani_regression_1996,chetverikov_cross-validated_2021}.

To evaluate the performance of the proposed method under sparse longitudinal data situation, in addition to generating $n_i$'s from $[16,31]$, we consider two alternative settings where $n_i$'s are randomly generated from the sets of integers in the intervals $[4,8]$ and $[6, 23]$ respectively while keeping other parameters unchanged. Table \ref{Tab_sparsity:ACC} summarizes clustering accuracies and mean squared errors under the three different settings for $n_i$'s. The proposed method performed reasonably well under sparse situations, and the performance improves as observations become denser.    


\begin{table}
  \centering
  \caption{Simulation results: average accuracy and the average mean squared error (MSE) of estimated mean functions.}
    \label{Tab_sparsity:ACC}
  \begin{tabular}{V{3} c|c|c|c V{3}}
    \hlineB{3}
            & \multicolumn{3}{c V{3}}{\textbf{With $\sigma^{2}_{non,k}$}}\\
    \cline{1-4} Intervals for $n_i$ & [4,8] & [6,23] & [16,31] \\
    \hlineB{2}
    Accuracy  & 0.6446 & 0.7354 & 0.8239 \\ 
    MSE $f_1$ & 0.0524 & 0.0394 & 0.0240 \\ 
    MSE $f_2$ & 0.0026 & 0.0014 & 0.0007 \\ 
    \hlineB{3}
    \end{tabular}
\end{table}

We included the codes in the supplementary materials and posted them to GitHub \\
(\url{https://github.com/HubDaniel/NMEM}). A subset of data is available at the NIH RADx Data Hub (\url{https://radx-hub.nih.gov/}).

\section{Conclusion and Future Work}

This article proposes a unified method for clustering longitudinal trajectories and relating the subgroups to other biological and clinical predictors. A flexible nonparametric mixed-effects mixture model is proposed to identify risk factors and classify patients with a change in body temperature before the diagnosis of COVID-19. We model the change in temperature using smoothing splines. We use penalized likelihood and the EM algorithm to estimate the mean functions, variance components, and covariates associated with the clustering probability. A simulation study shows that our method performs well and, under certain scenarios, outperforms existing methods. The results of data analysis suggest that different demographic characteristics influence the immune system response and provide an improved understanding of patient groups that may or may not experience a more severe course following infection with SARS-CoV-2.

    The identification of two clusters with respect to pre-diagnosis dynamics of body temperature is novel and significant. It indicates the existence of biological phenotypes that may react to infection differently during the COVID-19 incubation period. While the underlying biological reasons are unclear, it will be of interest in future analysis to explore the clinical course and patient outcomes after diagnosis of COVID-19 differ. One could hypothesize that patients with a pre-diagnostic rise in temperature may have a more severe infection (e.g., higher virus load) or augmented inflammatory response (e.g., higher levels of interleukin-6)  that may translate into worse COVID-19 outcomes. 
    
The identifiability of mixture models has received a great deal of attention, and some general conditions were provided in the existing literature 
\citep{teicher_identifiability_1963,holzmann_identifiability_2006,wang_note_2014,aragam_identifiability_2020,wong_semiparametric_2022}. 
However, these general conditions are complicated and difficult to verify. Specific conditions for the identifiability of the proposed model warrant further research.

\section*{Acknowledgement}
We would like to thank the editor and two anonymous reviewers for their insightful comments that significantly improved the manuscript.

\section*{Funding}
This research is partially supported by NIH grants R01-DK130067, R01-HL161303, R01-DK117208, and NSF grant DMS-2053423.

\appendix 
\section{Profiled Likelihood} \label{aP}
Recall equation (\ref{f1}) and (\ref{f2}) in the paper,
\begin{align} 
& \sum_{i=1}^{m}\left( w_{i 1} \log \left|\mathbf{V}_{i 1}\right| \right) + \left(\boldsymbol{y}-\mathbf{S} \boldsymbol{d}_{1}-\mathbf{R} \boldsymbol{c}_{1}\right)^{T} \mathbf{W}_{1}\left(\boldsymbol{y}-\mathbf{S} \boldsymbol{d}_{1}-\mathbf{R} \boldsymbol{c}_{1}\right)+N \lambda_{1} \boldsymbol{c}_{1}^{T} \mathbf{Q} \boldsymbol{c}_{1} \label{f3} \\
+ & \sum_{i=1}^{m}\left( w_{i 2} \log \left|\mathbf{V}_{i 2}\right| \right)+ \left(\boldsymbol{y}-\mathbf{S} \boldsymbol{d}_{2}-\mathbf{R} \boldsymbol{c}_{2}\right)^{T} \mathbf{W}_{2}\left(\boldsymbol{y}-\mathbf{S} \boldsymbol{d}_{2}-\mathbf{R} \boldsymbol{c}_{2}\right)+N \lambda_{2} \boldsymbol{c}_{2}^{T} \mathbf{Q} \boldsymbol{c}_{2}. \label{f4}
\end{align}

We can rewrite equations (\ref{f3}) and (\ref{f4}) as
\begin{align*}
\begin{split}
    & \sum_{i=1}^{m} \Big[ w_{i1} \log{|\mathbf{V}_{i1}|} + w_{i2} \log{|\mathbf{V}_{i2}|} + w_{i1} (\boldsymbol{y}_i-\mathbf{S}_i \boldsymbol{d}_1-\mathbf{R}_i \boldsymbol{c}_1)^T \mathbf{V}_{i1}^{-1} (\boldsymbol{y}_i-\mathbf{S}_i \boldsymbol{d}_1-\mathbf{R}_i \boldsymbol{c}_1) \\
    & + 
    w_{i2} (\boldsymbol{y}_i-\mathbf{S}_i \boldsymbol{d}_2-\mathbf{R}_i \boldsymbol{c}_2)^T \mathbf{V}_{i2}^{-1} (\boldsymbol{y}_i-\mathbf{S}_i \boldsymbol{d}_2-\mathbf{R}_i \boldsymbol{c}_2) \Big]
\end{split} \\
\begin{split}
    = & \sum_{i=1}^{m} \Big[n_i \log{\sigma^2}  + w_{i1}\log{|\mathbf{V}_{i1}^{*}|} + w_{i2}\log{|\mathbf{V}_{i2}^{*}|} \\
    & + \frac{w_{i1}}{\sigma^2} (\boldsymbol{y}_i-\mathbf{S}_i \boldsymbol{d}_1-\mathbf{R}_i \boldsymbol{c}_1)^T \mathbf{V}_{i1}^{*-1} (\boldsymbol{y}_i-\mathbf{S}_i \boldsymbol{d}_1-\mathbf{R}_i \boldsymbol{c}_1) \\
    & + \frac{w_{i2}}{\sigma^2} (\boldsymbol{y}_i-\mathbf{S}_i \boldsymbol{d}_2-\mathbf{R}_i \boldsymbol{c}_2)^T \mathbf{V}_{i2}^{*-1} (\boldsymbol{y}_i-\mathbf{S}_i \boldsymbol{d}_2-\mathbf{R}_i \boldsymbol{c}_2) \Big], 
\end{split}
\end{align*}
where $\mathbf{V}_{i1}^{*} = \mathbf{V}_{i1}/\sigma^2$ and $\mathbf{V}_{i2}^{*} = \mathbf{V}_{i2}/\sigma^2$. 

Therefore we have,
\begin{align*}
\begin{split}
\hat{\sigma}^2 = \frac{1}{N} \sum_{i=1}^{m} \Big[ 
&w_{i1} (\boldsymbol{y}_i-\mathbf{S}_i \boldsymbol{d}_1-\mathbf{R}_i \boldsymbol{c}_1)^T \mathbf{V}_{i1}^{*-1} (\boldsymbol{y}_i-\mathbf{S}_i \boldsymbol{d}_1-\mathbf{R}_i \boldsymbol{c}_1) + \\
&w_{i2} (\boldsymbol{y}_i-\mathbf{S}_i \boldsymbol{d}_2-\mathbf{R}_i \boldsymbol{c}_2)^T \mathbf{V}_{i2}^{*-1} (\boldsymbol{y}_i-\mathbf{S}_i \boldsymbol{d}_2-\mathbf{R}_i \boldsymbol{c}_2) \Big].
\end{split}
\end{align*}

Plugging the above back we get the profiled likelihood:
\begin{align*}
l_{p} = N \log \Big(\sum _ { i = 1 } ^ { m } \Big[
&w_{i 1}\left(\boldsymbol{y}_{i}-\mathbf{S}_{i} \boldsymbol{d}_{1}-\mathbf{R}_{i} \boldsymbol{c}_{1}\right)^{T} \mathbf{V}_{i 1}^{*-1}\left(\boldsymbol{y}_{i}-\mathbf{S}_{i} \boldsymbol{d}_{1}-\mathbf{R}_{i} \boldsymbol{c}_{1}\right)+ \\
&w_{i 2}\left(\boldsymbol{y}_{i}-\mathbf{S}_{i} \boldsymbol{d}_{2}-\mathbf{R}_{i} \boldsymbol{c}_{2}\right)^{T} \mathbf{V}_{i 2}^{*-1}\left(\boldsymbol{y}_{i}-\mathbf{S}_{i} \boldsymbol{d}_{2}-\mathbf{R}_{i} \boldsymbol{c}_{2}\right)\Big]\Big) \\
&+ \sum_{i=1}^{m}\Big[w_{i 1} \log \left|\mathbf{V}_{i 1}^{*}\right|+w_{i 2} \log \left|\mathbf{V}_{i 2}^{*}\right|\Big].
\end{align*}

\bibliographystyle{unsrtnat}
\bibliography{references}  






\end{document}